\documentclass{sigplanconf}

\usepackage{amsmath}
\usepackage{amsthm}

\usepackage{algorithm2e}

\newtheorem{definition}{Definition}

\newcommand{\Stmt}{\Omega} 
\newcommand{\Field}{{\cal F}}
\newcommand{\Proc}{{\cal P}}
\newcommand{\Dom}{{\cal D}}
\newcommand{\Idx}{{\cal I}}
\newcommand{\Scat}{{\cal T}}

\newcommand{\dquote}[1]{``#1''}

\begin{document}

\setlength{\pdfpageheight}{\paperheight}
\setlength{\pdfpagewidth}{\paperwidth}

\conferenceinfo{PLDI '14}{June 9--11, 2014, Edinburgh, United Kingdom} 
\copyrightyear{2014} 
\copyrightdata{978-1-nnnn-nnnn-n/yy/mm} 
\doi{nnnnnnn.nnnnnnn}





\title{Introducing Molly: Distributed Memory Parallelization with LLVM}


\authorinfo{}
           {}
           {}

\maketitle

\begin{abstract}
Programming for distributed memory machines has always been a tedious task, but necessary because compilers have not been sufficiently able to optimize for such machines themselves. Molly is an extension to the LLVM compiler toolchain that is able to distribute and reorganize workload and data if the program is organized in statically determined loop control-flows. These are represented as polyhedral integer-point sets that allow program transformations applied on them. Memory distribution and layout can be declared by the programmer as needed and the necessary asynchronous MPI communication is generated automatically. The primary motivation is to run Lattice QCD simulations on IBM Blue Gene/Q supercomputers, but since the implementation is not yet completed, this paper shows the capabilities on Conway's Game of Life.
\end{abstract}




\section{Introduction}\label{sct:introduction}

Since the standardization of MPI by far most scientific programs on distributed memory machines are written using this API. In order to exploit today's shared memory multicore architectures the most performant programs are hybrid MPI/OpenMP with the cost of more complicated programming and program maintenance. Even worse, different hardware platforms require to completely reorganize the program for optimal performance. For instance, an architecture may have a network with high bandwidth but relative slow CPUs whereas on a second platform it is the other way around. Optimization for the second may do the same computation on the receiving and sending node just to avoid saturating the network.

Therefore, writing fast programs may require rewriting the performance-sensitive parts for every architecture. Which choice of optimization is the fastest on a particular machine can usually not be known in advance. Scientific applications exists for the results they deliver, not for the joy of optimizing them. Hence, although not very successful in the past, the quest for compiler-driven data distribution remains worthwhile. Even though no compiler knows the optimal code for all the cases, it can assist the programmer by heuristically generating the missing parts once the programmer chooses the performance-relevant ones.

This work has been driven from the experience of manually optimizing a program that does Lattice QCD simulation~\cite{tmlqcd,bonati13} for the IBM Blue Gene/Q supercomputer. The data layout was changed multiple times in order to find out which the fastest one is. Every time large portions had to be rewritten such that the overhead remains low for a fair comparison. The different versions were conceptually similar, which motivated to let the compiler do the main work using already existing techniques that represent a program as a polyhedron~\cite{fautrier92a,fautrier92b,darte00}. 

The core of most Lattice QCD solvers is an 8-point stencil in 4 dimensions with 1320 floating-point operations and 2688 bytes to read per site update in double precision. It is the most optimized part and any overhead may significantly slow down the solver compared to hand-optimized code. A compiler-optimized version has to be similar to the equivalent manually optimized code in order to be useful or physicists will continue optimizing code by hand. Dynamic techniques at runtime such as task scheduling are therefore off the table. 

Usually just the instructions are represented and reordered in the polyhedral model, although also the elements of an array can be modeled this way. The alpha language does this to compact array storage~\cite{mauras90,quillere02}. It is however rarely used to just change the location of data elements although it is a powerful representation. \cite{grosslinger09}~suggests this for scratchpad memories, this paper for homogeneous distributed memory machines (DMM).

This work extends the Clang+LLVM~\cite{llvm,clang} compiler toolchain with a new pass called \emph{Molly}. The choice fell to LLVM because (1) it is open source, (2) relatively easily extendable, (3) promotes link-time optimization (4) it already contains many necessary components, especially Polly and (5) can act as a Just-In-Time (JIT) compiler.

Point (3) is important because reordering storage is an inherently global process. Different components cannot be translated inependently if they do not agree on the data layout of process-wide data. Polly~\cite{polly} is the polyhedral optimization subproject of LLVM. It provides detection of static control parts (SCoPs), is able to optimize them and generates code again. Molly intercepts Polly to modify the SCoPs before code re-generation.

The fifth point is relevant because the current implementation of Molly requires the size of distributed arrays and the geometry of compute nodes to be known at compile time. This is because block-cyclic (or just any interesting) data distribution is not a linear function. If $N$ is the length of an array, $P$ the number of processors, $i$ an element of the array array, $p$ a processor number, and $l$ the index of element $i$ on processor $i$, then the relation for block-distribution is $i = p(N/P) + l$. The term $p(N/P)$ is not of the form \dquote{constant times variable} required to be an affine term. $p$ naturally varies in an MPI/SPMD program, therefore $N/P$, the block size, must be constant. Future versions may precompute $p(N/P)$ and parametrize the SCoP using this newly introduced variable, such that at least block-distributions do not need to predefined the geometry of the cluster machine.

The \emph{Integer Set Library} (ISL)~\cite{isl} is, despite its name, a library for sets of vectors of integers and rationals represented as unions of $\mathbf{Z}$-polyhedra. It is already used by Polly and Molly uses it as well for indexsets.

This article is organized as follows. Section~\ref{sct:related} gives an overview on other solutions on the same problem. Section~\ref{sct:toolchain} introduces the toolchain and what parts had to be modified for Molly. In Section~\ref{sct:communication} the main part about how generating communication code is presented and Section~\ref{sct:conways} shows this on the example of Conway's Game of Life. Experimental execution results are presented in section~\ref{sct:experiments} and possible further developments discussed on Section~\ref{sct:extensions}. Finally, Section~\ref{sct:conclusion} summarizes this work.

\section{Related Work}\label{sct:related}

In the past, efforts to shift the responsibility of parallelization and data distribution to the compiler were not very successful. Probably the most noteworthy project to relief the programmer of the burden of explicit parallelization is High-Performance Fortran (HPF). Today, HPF plays a  minor role because of many reasons~\cite{kennedy07}. One of them is that compilers promised to optimize code, but could only do so for specifically structured code. Every compiler could only optimize a different set of code patterns.

Probably the most similar work is~\cite{adve98}. The authors also build polyhedral subsets of array indexes that have to be transferred to different nodes. They call a \dquote{communication event} is called a chunk in this paper. However, their set operations are very different and also does more advanced techniques like index set splitting and merge of communication events. Their tool dHPF is a HPF to Fortran77 with MPI source-to-source compiler and uses the Omega~\cite{omega} library for the set operations. It only allows block-cyclic data distribution and uses the owner-computes policy for statements. More work of this kind are the Last Write Tree~\cite{amarasinghe93} and \cite{dathathri13}.

A different kind of optimization partitions the work into tiles which are then distributed to the nodes~\cite{bondhugula11,classen06,yuki13}. The tiles size is fixed at compile-time, but the number of nodes doesn't need to be fixed until runtime. Therefore most of the work uses a notion of \dquote{virtual} processors. Virtual processors are mapped to physical ones at runtime.

A novel technique has been presented in~\cite{ravishankar12}. It combines syntactical code analysis with a dynamic part that allows non-static code behavior such as indirect array accesses. These can normally not be represented in a SCoP.

Again another approach is to invent a language that has high-level operators on multi-dimensional arrays. ZPL~\cite{snyder99} belongs to this category. Instead of \dquote{accessing the element to the left} there is a shift operator that moves the entire field one index to the right. Such languages require rethinking of algorithms, but high-level operators can be pre-implemented and algebraically optimized efficiently.

The middle way are extensions to existing languages. To name a few, there is OpenACC, OmpSs, C++ AMP, Universal Parallel C (UPC), etc. The first three are language extensions that targeting accelerators like GPUs to offload computation from the main CPU. The programmer has to state explicitly the data region that must be available on the device memory. UPC, a representative of languages using PGAS (Partitioned Global Address Space), create an address space window for non-local data. Code flow still has to be explicitly organized and synchronized.

\section{The Toolchain}\label{sct:toolchain}

The reduced variant of Conway's Game of Life shown in Figure~\ref{fig:conway} illustrates a program that Molly can optimize. The original Game of Life is a 9-point stencils whereas this variant has only 5 points. The example does 100 iterations and copies back the last iteration's result because pointer swaps cannot (yet) be handled in SCoPs.

\begin{figure}[htb]
{\small\begin{verbatim}
#include <molly.h>

[[molly::pure]] 
bool hasLife(bool hadLife, int neighbors) {
  if (hadLife)
    return 2 <= neighbors && neighbors <= 3;
  return neighbors == 3;
}

molly::array<bool,1024,1024> front;
molly::array<bool,1024,1024> back;

int main() {
  for (int i = 0; i < 100; i+=1) {
    for (int x=1,w=front.length(0); x < w-1; x+=1)
      for (int y=1,h=front.length(1); y < h-1; y+=1) {
S1:      auto neighbors = front[x-1][y] + front[x][y+1];
         neighbors += front[x+1][y] + front[x][y-1];  
         auto living = hasLife(front[x][y], neighbors);     
         back[x][y] = living;
      }
    for (int x=1,w=back.length(0); x < w-1; x+=1)
      for (int y=1,h=back.length(1); y < h-1; y+=1)
S2:     front[x][y] = back[x][y];
  }
  return EXIT_SUCCESS;
}
\end{verbatim}}
\caption{Conway's Game of Life}\label{fig:conway}
\end{figure}

Such \emph{molly::array}-typed variables that are distributed are called \emph{fields} in this paper. This comes from physics' jargong of a quantity for every point in spacetime. They are a typical application for such arrays, notably in lattice QCD.

\subsection{The Clang Part}

Clang~\cite{clang} is the C, C++ and Objective-C frontend that compiles source files of those languages into LLVM~\cite{llvm} intermediate representation, also called LLVM IR. It has to be modified in order to pass information on distributed arrays to the optimizer as metadata.

First, there is a new header file \texttt{molly.h} which declares the \texttt{molly::array} variadic template class. It represents a distributed array of arbitrary size and dimensions. Its semantics are different than the standard C++ arrays. Pointer arithmetic beyond the current element are invalid. The compiler may change the elements' order, the value might be outdated or even not stored on the local node.

The header file uses an overloaded index operator in order to make field accesses like \texttt{back[x][y] = living} resemble

\begin{verbatim}
bool *ptr = __builtin_molly_ptr(&back, x, y);
*ptr = living;
\end{verbatim}

\texttt{\_\_builtin\_molly\_ptr} is a compiler-builtin and returns a pointer. The value of the pointer is meaningless if the value has no location on the current node. Later passes need to fix-up for this.
  
Clang also generates metadata for every distributed array. The information include what the element type is, as well as the shape of the array. The template arguments are not kept when translated to the respective LLVM type. 

In principle, any language frontend can generate such kind of accesses and forward it to LLVM, but for the current project only C++ is supported.

\subsection{The LLVM Part}

The intermediate representation of the previous snippet that is passed to LLVM is shown in Figure~\ref{fig:mollyptr}. The newly introduced LLVM intrinsic \texttt{llvm.molly.ptr} semantically resembles the \texttt{GetElementPtr} instruction: it returns the pointer to an array's element, but logically allows accessing remote memory.

\begin{figure*}
\begin{verbatim}
  %1 = call i8* llvm.molly.ptr(%"class.molly::array"* @back, i64 %x, i64 %y) 
  store i8 %living, i8* %1
\end{verbatim}
\caption{Equivalent LLVM IR code of \dquote{\texttt{back[x][y] = living}}}\label{fig:mollyptr}
\end{figure*}

Because pointer arithmetic is disallowed on values returned by \texttt{llvm.molly.ptr}, only \texttt{load} and \texttt{store} instructions can use it. Whenever the Molly pass encounters a load/store of a pointer returned by \texttt{llvm.molly.ptr} it may look up the accessed field and the coordinates to know what element is accessed.

Molly is a module-level pass, therefore it has access to all \texttt{molly::array<>} variables and all the location they are used. Hence, it is possible to align value distributions of the fields, but is currently not implemented. A simple block-distribution is used and instructions are executed on the node that owns the value they compute (owner computes rule).

The next step is to create an isolated basic block for every field access. The splits must happen because the data source domains where either local data of remote data is accessed do not match. For instance, the data for \texttt{front[x-1][y]} is local to node $(0,0)$ on iteration $(x,y)=(512,1)$, but on node $(1,0)$ for the access \texttt{front[x][y+1]} although both accesses must be executed on the same node because they both access the scalar \texttt{neighbors}. 

Data between the statement is exchanged through variables on the stack, for instance the variable \texttt{neighbors}. There is no inter-node transfer generated for such non-field scalar variables and therefore must always be executed on the same node.

\subsection{Polly: SCoP-Detection}

At this stage Polly detects whether a loop structure is SCoP-compatible and if so, identifies the statements and their natural scheduling parameters (iteration domain and relative ordering). The isolation from the previous step ensures that every field access gets its own statement and can therefore scheduled independent of the rest.

Next, Molly introduces two virtual statements that are executed before and after all the statements in the SCoP, called prologue and epilogue. The prologue virtually writes everything and the epilogue reads every data location referred to in the SCoP. They are use to compute the data flow into and out of the SCoP.

The data dependencies between the statements are computed by ISL. Molly requires the flow dependencies (read-after-write) of data in fields and local memory independently. Other types of dependencies (write-after-write and write-after-read) are irrelevant because Molly does not change the ordering of instructions, but needs to know where data is used and generated.

ISL determines direct dependencies only, i.e. for every value that is read it is able to determine the unique statement instance that computed that value, as long as all array index expressions are affine.

\subsection{Molly}

Then, Molly has to decide on which node a statement is executed. By default, statements are executed preferably on the same node as one of the statements that produces its input. Statements that produce data that needs to be written back at the SCoP's end (i.e. has a dependency to the epilogue) is computed on that data's home location. Analogous for statements that read data from before the statements, i.e. dependencies from the prologue.

Statements that produce data written to a node-local variable (i.e. non-field flow dependency) need to be executed on at least all the nodes that potentially consume it. 

The next section also belongs to the Molly-part, but explains the core message passing optimization and therefore has its own section.

\section{Communication Code Generation}\label{sct:communication}

The main part is the generation of code that transfers the data between the nodes. We start off explaining some terminology used in polyhedral optimization and notation used in this paper.

\subsection{Application Modeling}

Molly needs information about six different spaces and the relation between them in order to transform a program. The spaces are

\begin{itemize}
\item $\Stmt$, the set of all statement in the SCoP. Typical names for statements are $R$, $S1$, $S2$, $G$, $C$ etc.
\item $\Proc \subseteq \mathbf{Z}^{N_\Proc}$, the set of computing nodes or processors; also called the cluster space. A coordinate is typically named $(a,b)$ or $\vec p$.
\item $\Dom_S \subseteq \mathbf{Z}^{N_S}$, the domain of the statement $S \in \Stmt$; also called the iteration space. 
Typical variable names are $\vec i$, $\vec j$. $(i, j)$ can also denote a single coordinate.
\item $\Field$, the set of fields. The only specific fields on this paper are $front$ and $back$, the arrays from Figure~\ref{fig:conway}.
\item $\Idx_e \subseteq \mathbf{Z}^{N_e}$, the set of addressable indexes of a field $e \in \Field$, therefore called the indexset space. An index is typically named $\vec k$.
\item $\Scat \subseteq \mathbf{Z}^{N_\Scat}$, the scheduled time of the statements in the current SCoP, also called the scatter space.
\end{itemize}

The $N_{\Proc,S,e,\Scat}$ describe the number of dimensions of the spaces. A tuple $(S, \vec i) \subseteq \Stmt \times \Dom_S$ is called an \emph{instance} of statement $S$. A tuple $(S, \vec i, \vec p) \subseteq \Stmt \times \Dom_S \times \Proc$ is called an \emph{execution} of instance $(S, \vec i)$ on node $\vec p$. 

The previous analyses of a SCoP by Polly and Molly resulted in the following functions and relations between those spaces:

\begin{itemize}
\item $\theta_S: \Dom_S \rightarrow \Scat$, the schedule of every statement representing the time of execution in a sequential program.
\item $\lambda_S: \Dom_S \rightarrow 2^{\Field \times \Idx}$, the field elements a statement accesses, usually at most one, with the exception of the prologue and epilogue.
\item $(G, \vec i_G) \,\delta\, (C, \vec i_C)$ is the flow-dependence relation in which the consumer instance $(C, \vec i_C)$ reads a value computed by the producer/generator instance $(G, \vec i_G)$.

\item $\pi_{e \in \Field}: \Idx_e \rightarrow 2^\Proc$, the node that owns and stores a field element, called the home location. Multiple nodes may be responsible for the same element in which case all of them must store the same value.
\item $\pi_{S \in \Stmt}: \Dom_S \rightarrow 2^\Proc$, on which node a statement will be executed. A single statement can also be executed on several nodes. This is called the \emph{where}-mapping. An execution $(S, \vec i, \vec p)$ is therefore restricted to the nodes $\vec p \in \pi_S$.
\end{itemize}

For any two statements $G$ and $C$ let $(G,i_G)\; \delta \;(C,i_C)$ the data flow dependence between a data-generating statement $G$ and a consumer $C$. They both access the field element $\lambda_G(\vec i_G) = \lambda_C(\vec i_C)$. In case the statement instances are executed on different nodes, some communication code must be generated between them.

To avoid transferring single values with immense overhead values are grouped into \emph{chunks}, such that all data of one chunk can be written to the same buffer (per destination), then sent and received. Two values cannot be in the same chunk iff there is a direct or indirect dependency between the statement that produce them.

\begin{definition}[Chunking Function]
A \emph{chunking function} is an projection $\varphi: (\Stmt, \Dom) \rightarrow (\Stmt, \Dom)$ of statement instances to representative instances. By convention $\varphi$ is idempotent, i.e. any representative instance projects to itself. The equivalence classes $[\varphi(R, \vec j)] = \{ (S, \vec i) \mid \varphi(R, \vec j) = (S, \vec i) \}$ are called \emph{chunks}.
\end{definition}

A chunking function is a mean to identify chunks. One can think of the representative instance as the first statement of the chunk, although an order may not have been established yet. In fact, any image space can be used, but any other choice would be more arbitrary.

\begin{definition}[Valid Chunking Function]
Let $\delta^*$ be the transitive closure of flow dependencies between all statements. A chunking function $\varphi$ is \emph{valid} if it does not violate any dependencies in $\delta^*$, i.e. $\not\exists (S,\vec i,S,\vec i) \in \varphi(\delta^*)$, given that $\delta^*$ was irreflexive before.
\end{definition}

The objective is to find a valid chunking function with as few elements as possible and preferably whose chunks are about the same size. The worst chunking, but always possible function is the identity function. This yields to every value transferred individually as mentioned before.

If the chunking function as well as the dependencies are (piecewise) affine, finding a chunking function with minimal chunks involves finding a solution of an inequality system.

Let $(G, \vec i_G) \rightarrow (C, \vec i_C)$ be a dependency of $\delta^*$. $\vec i_G \neq \vec i_C$ because otherwise the instance requires data that itself produces which is invalid. Let $\varphi_{\{G,C\}}(\vec i) = \alpha_0 i_o + \alpha_1 i_1 \dots$ a possible chunking function with unknown $\alpha_{0, \dots}$. For validity, we have to find $\alpha_{0,\dots}$, s.t. $\varphi(\vec i_G) \neq \varphi(\vec i_C)$ for all flow dependencies. The goal is a function $\varphi$ whose image spans minimally per dimension.

This problem is analogous to the scheduling problem to find a schedule with minimal span and therefore maximal parallelism. One can think of the chunking function collapsing a set of statement instances that are executed in parallel, but here their data is received at the same time. An algorithm is suggested in~\cite{bastoul03}.

Molly uses a much simpler algorithm which exploits that loops are typically already written in a way that allow chunking of innermost loops. The algorithm finds the outermost loop such that (1) all consumer instances of one chunk are executed after all of the generator instances and (2) the resulting chunking function is valid. (2) implies (1) because the consumer is always dependent on the generator and therefore is always scheduled later, but it is much faster to check and covers most practical cases. Also, this gives the direct scatter locations where to insert the send and receive statements and therefore guarantees schedulability. 
The algorithm is shown as Algorithm~\ref{alg:chunking}.

\SetKw{KwContinue}{continue}
\SetKw{KwLambda}{lambda}
\SetKw{KwNot}{not}

\begin{algorithm}
\SetAlgoLined
\SetKwInOut{Input}{Input}
  \Input{parametric flow dependence $(G, \vec i_G) \;\delta\; (C, \vec i_C)$}
  \KwResult{chunking function $\varphi$}

  \For{$l$=0 (entire SCoP) \KwTo $N_\Scat - 1$ (innermost loop of $C$)}{
    \lIf(\tcp*{(1)}){\KwNot $\theta(G, \vec i)_{0, \dots, l} <_\text{lex} \theta(C, \vec i)_{0, \dots, l}$}{\KwContinue} 
    $\varphi$ = \KwLambda: $(C, \vec i) \mapsto (C, (i_0, i_1, \dots, i_l, 0, \dots, 0))$ \;
    $\delta'$ := apply $rho$ on any instance $(C, \vec i_C)$ on the left or right side of $\delta$ (collapse) \;
	\lIf(\tcp*{(2)}){ $\varphi(C, \vec i_C) \;\delta'^*\; \varphi(C, \vec i_C) $ }{\KwContinue}
	    
    \Return $\varphi$ \;
  }
   \Return \KwLambda: $(C, \vec i) \mapsto (C, \vec i)$ \;
   
  \caption{Chunking function heuristic}\label{alg:chunking}
\end{algorithm}

Once the chunking function id known, Molly can insert calls that trigger communication and redirect any accesses to remote data to the communication buffers.

First, Molly builds a giant relation between the representative instance of the chunk, generator execution, consumer execution, the field element they access:

\begin{align*}
  Transfers := \big\{ ((R, \vec i_R), (G, \vec i_g, \vec p_g), (C, \vec i_C, \vec p_C), (F, \vec k_f)) \mid \\
     \vec i_g \in \Dom_G, \vec p_g \in \pi_G(i_g), \\
     \vec i_c \in \Dom_C, \vec p_c \in \pi_C(i_g),\\
     (G, \vec i_g) \delta (C, \vec i_c),\\
     \lambda_C(G, \vec i_g) = (F, \vec k_f) = \lambda_C(\vec i_g),\\
     (R, \vec i_R) = \varphi_C(\vec i_C) 
\big\}
\end{align*}

A generator instance can have multiple executions on different nodes and therefore one has to be picked, preferably the one that executed on the same node as the consumer if there is one. Molly's algorithm for this is to pick the lexicographical minimal. The result, $Transfers'$, has a unique tuple $(G, \vec i_g, \vec p_g)$ for every tuple $(C, \vec i_C, \vec p_C), (F, \vec k_f)$, and therefore can be defined as a function $Source: (C, \vec i_C, \vec p_C, F, \vec k_f) \mapsto (G, \vec i_g, \vec p_g)$.

To construct the communication events, the relation elements in $Transfers'$ are grouped by the chunk they belong to (Equation~\ref{eq:chunky}).

\begin{figure*}[tb]
\begin{align}
  Chunk&: (R, \vec i_R) \mapsto  \big\{  ((G, \vec i_g, \vec p_g), (C, \vec i_C, \vec p_C), (F, \vec k_f)) \big | ((R, \vec i_R), (G, \vec i_g, \vec p_g), (C, \vec i_C, \vec p_C), (F, \vec k_f)) \in Transfers' \big\} \notag \\
  Chunks &:= \big\{ Chunk(R, \vec i_R) \,\big|\, Chunk(R, \vec i_R) \neq \emptyset \big\} \label{eq:chunky}
  \end{align}
  \begin{align}
    \theta_\text{send\_wait}(R, \vec i_R) &= \min_\text{lex}( \{ \ \theta(G, \vec i_G) | ((G, \vec i_g, \vec p_g), (C, \vec i_C, \vec p_C), (F, \vec k_f)) \in Transfers' \}) - ( 0\dots , 1 ) \label{eq:beforegenerator} \\
    \theta_\text{recv\_wait}(R, \vec i_R) &= \min_\text{lex}( \{ \ \theta(C, \vec i_C) | ((G, \vec i_g, \vec p_g), (C, \vec i_C, \vec p_C), (F, \vec k_f)) \in Transfers' \}) - ( 0\dots , 1 ) \label{eq:aftergenerator} \\
    \theta_\text{recv}(R, \vec i_R) &= \max_\text{lex}( \{ \ \theta(C, \vec i_C) | ((G, \vec i_g, \vec p_g), (C, \vec i_C, \vec p_C), (F, \vec k_f)) \in Transfers' \}) + ( 0\dots , 1 ) \label{eq:afterconsumer}
\end{align}
\end{figure*}

For any such chunk Molly creates two static variables, one for the sending send and another for the receiving side, and six statements which replace the consumer executions.

\begin{enumerate}
\item A call to \texttt{\_\_molly\_send\_wait}, which waits for the previous chunk to finish sending data if still active and returns a pointer to the memory buffer into which the data of the current chunk must be written. The scatter function on node $\vec p_G$ to node $\vec p_C$ is Equation~\ref{eq:beforegenerator}, i.e. one timestep before the first generator execution. To avoid overlapping with an already existing instance at this timestep one can simply multiply all the scatter function by 2 or more beforehand.

\item A new generator statement. Instead of writing to local memory, it writes the buffer returned by the previous statement at an index that is computed from the hull of field indices that are to be transferred between the sending and receiving node. The scatter function is the same as the original generator statement.

\item A call to \texttt{\_\_molly\_send} that starts the transfer. Its scatter is one step after the last generator execution of this chunk.

\item A call to \texttt{\_\_molly\_recv\_wait} on every receiving node $\vec i_C$. Its scatter function is Equation~\ref{eq:aftergenerator}, one step before the first consume execution on this node. This runtime function returns a pointer to the buffer that has been received.

\item A replacement consumer statement which delegates the original return value of \texttt{\_\_molly\_ptr} to the receiving buffer at the index computed the same way as the sendbuffer of the generator statement.

\item A call to \texttt{\_\_molly\_recv} that releases the receive-buffer and resets the communication object in order to be prepared for the next chunk. Its scatter function on node is Equation~\ref{eq:aftergenerator}, that is, one timestep behind the last replacement consumer.

\end{enumerate} 

In addition, Molly generates calls initialization- and release-functions for every chunk and transfer source/destination 
. They setup (respectively free) the target node coordinate, the buffer size and an unique identifier (\emph{tag}) for every communication event in advance. For the receiving nodes, it also resets the buffer to ready state.

Finally, all the original generator statements are removed. All the consumers now read from the communication buffers such that the original writes are unused. As a result, no calls to \texttt{llvm.molly.ptr} remain active in this SCoP.

Dataflows from the prologue and to the epilogue are special cases of the more general flow between arbitrary statements. The difference is that the reading part reads from, respectively writes to the local storage of the field access per node. The prologue and epilogue are singleton statements (per node) that touch all the data. Therefore, there is just a single chunk that transfers data to, respectively out of the SCoP. Flows from the prologure to the epilogue can be ignored.

Since Polly does not understand placements of statement instances ($\pi$) one has to reduce them to something more generic. The coordinate of the currently executing node is added as a parameter of the SCoP (the same way runtime parameters like the upper bound of a for-loop). Then, the iteration relation $\vec i$, which implicitly uses these parameters, is modified such that the statement instances are removed that

\subsection{Polly: Code Optimization and Generation}

Once the communication statements and iteration domains are fixed, the set of statements are given to Polly again. Polly may apply additional optimizations that reorder statements by applying different scatter functions ($\theta$), vectorize and parallelize using OpenMP. There are two different optimization engines that Polly supports: PLuTo~\cite{pluto} and ISL~\cite{isl}.

Polly's final part is to generate IR code again from the statement instances. Again, Polly supports two engines: Cloog~\cite{cloog} and ISL~\cite{isl}.  The original code will be deactivated and finally removed by the dead code elimination pass.

ISL is required by Polly in any case and can do optimization and code generation, the other two engines are therefore optional.

\subsection{Finalization}

A few smaller task remain to be done. There can be leftover \texttt{llvm.molly.ptr} remaining in the code that were not inside a SCoP because they potentially access non-local data and they are unknown to the LLVM machine code generator\footnote{Unsurprisingly, there is no assembly instruction that can localize remote memory on DMMs. That's why they are termed that way.}. Accesses to its returned pointer are replaced by the runtime functions with the names \texttt{\_\_molly\_value\_load} and \texttt{\_\_molly\_value\_store} that handle single value transfers in a synchronous manner. This is slow, but necessary for correctness.

Moreover, a set of functions for accessing the local data storage is being generated. The data ordering is just known to the compiler, therefore the runtime needs a way to find the correct index inside local memory. They are needed, for instance, to implement the previous functions that transfer single values between nodes.

Finally the \texttt{main} function is wrapped by another runtime function which now serves as the entry point. The runtime main initializes the communication API and calls the transfer buffer initialization functions, calls the application's \texttt{main}, and releases any resources before exiting the application. As additional parameters it receives the geometry of nodes the application has been compiled for. It must abort if it doesn't match the geometry it is running an since correctness is not guaranteed anymore.

The runtime needs to call the application's \texttt{main} function by itself because some communication APIs may change the application's command line parameters. Noteably \texttt{MPI\_Init} is allowed to do this if the implementation submits the command line parameters by some other way than by calling the executable on the target node with said parameters in advance.

\subsection{MollyRT}

MollyRT is the runtime library for applications compiled with Molly. It provides implementations of functions that are called during the code generation as mentioned in the previous section.

The current library's only backend is the Message Passing Interface (MPI) which is available on virtually every cluster supercomputer these days. But other interfaces are thought of as well, especially the close-to-the-metal System Programming Interface (SPI) of the Blue Gene/Q. The MPI overhead on this machine is immense. Other possible backends are possible of as well: Transfers between threads/processes of a shared memory machine, SHMEM, IBM PAMI, PGAS, etc.

The mapping between the MollyRT functions and the MPI functions is obvious in ost cases. Communication buffers setup requests using MPI persistent transfers upon their initialization. The \texttt{\_wait}-functions call \texttt{MPI\_Wait} on these. \texttt{\_\_molly\_send} and \texttt{\_\_molly\_recv} issue an \texttt{MPI\_Start}.

\section{Example: Conway's Game of Life}\label{sct:conways}

The first part of the Molly passes is the isolation of basic blocks that contain field accesses. Originally, the Game of Life code in Figure~\ref{fig:conway} has two blocks inside the loops that do not belong to a for-loop. We can ignore the return statement as is does not contain a field access nor does it belong to a SCoP. Isolation yields 9 basic blocks resembling Figure~\ref{fig:isolated}.

\begin{figure}[htb]
{\small
\begin{verbatim}
  for (int i = 0; i < 100; i+=1) {
    for (int x=1, w=front.length(0); x < w-1; x+=1)
      for (int y=1, h=front.length(1); y < h-1; y+=1) {
S1.1:   int neighbors = front[x-1][y] 

S1.2:   neighbors += front[x][y+1];
         
S1.3:   neighbors += front[x+1][y];
        
S1.4:   neighbors += front[x][y-1];  
         
S1.5:   auto hadLife = front[x][y];
         
S1.6:   auto living = hasLife(hadLife, neighbors);  
            
S1.7:   back[x][y] = living;       
      }
    for (int x=1, w=back.length(0); x < w-1; x+=1)
      for (int y=1, h=back.length(1); y < h-1; y+=1) {
S2.1:   auto tmp = back[x][y];
      
S2.2:   front[x][y] = tmp;
      }
  }
\end{verbatim}
}
\caption{SCoP with isolated field accesses}\label{fig:isolated}
\end{figure}

Running Polly's SCoP-detection then creates a SCoP and individual statements from these basic blocks. The analysis result below shows the statements with their domains ($\Dom$) and scatter function $\theta$. The scatter space $\Scat$ is determines by the number of loops (=3) plus the order of statements within the loops, respectively the order of the outermost loops and statements.

\begin{align*}
  \Stmt &= \{ S1.1, S1.2, S1.3, S1.4, S1.5, S1.6, S1.7, S2.1, S2.2 \}\\
  \Dom_{S1.1 \dots S1.7} &= \{ (i, x, y) \mid 0 \le i < 100, 1 \le x,y < 1023 \}\\
  \Dom_{S2.1, S2.2} &= \{ (i, x, y) \mid 0 \le i < 100, 1 \le x,y < 1023 \}\\
  \Scat &= {\cal Z}^7\\
  \theta_{S1.\alpha}(i,x,y) &= (0, i, 0, x, 0, y, \alpha)\\
  \theta_{S2.\alpha}(i,x,y) &= (0, i, 1, x, 0, y, \alpha)
\end{align*}

Molly reads the metadata generated by Clang to find which global variables are fields. In this example, there are two fields: $front$ and $back$. The indexset domains are always zero-based and the dimension's sizes have been declared in the source files. The fields are accessed in the isolated statements with the elements accessed determined by the $\lambda$-function.

\begin{align*}
  \Field &= \{ front, back \}\\
  \Idx_{front,back} &= \{ (w,h) \mid 0 \le w,h < 1024 \}\\
  \lambda_{S1.1}(i,x,y) &= \{ (front, (x-1,y)) \}\\
  \lambda_{S1.2}(i,x,y) &= \{ (front, (x,y+1)) \}\\
  \lambda_{S1.3}(i,x,y) &= \{ (front, (x+1,y)) \}\\
  \lambda_{S1.4}(i,x,y) &= \{ (front, (x,y-1)) \}\\
  \lambda_{S1.5}(i,x,y) &= \{ (front, (x,y))   \}\\
  \lambda_{S1.6}(i,x,y) & = \emptyset \\
  \lambda_{S1.7}(i,x,y) &= \{ (back, (x,y))   \}\\
  \lambda_{S2.1}(i,x,y) &= \{ (back, (x,y))   \}\\
  \lambda_{S2.2}(i,x,y) &= \{ (front, (x,y))   \}
\end{align*}

To compute the flow a data into and out of the SCoP, the virtual statements $Prologue$ and $Epilogue$ are added to the list of statements. They execute just once (Equation~\ref{eq:logueonce}), upon entering the SCoP (Equation~\ref{eq:prologuebefore}), respectively after the execution of all statements (Equation~\ref{eq:epilogueafter}). The both touch all the data, the prologue is a virtual write access and the epilogue as a read access.

\begin{align}
  \Stmt' &= \Stmt \cup \{ Prologue, Epilogue \} \notag\\
  \Dom_{Prologue,Epilogue} &= \{ () \} \label{eq:logueonce}\\
  \theta_{Prologue}() &= (-1,0,0,0,0,0,0) \label{eq:prologuebefore}\\
  \theta_{Epilogue}() &= (+1,0,0,0,0,0,0) \label{eq:epilogueafter}\\
  \lambda_{Prologue,Epilogue}() &= \left\{ (e,(w,h)) \;\middle|\; e \in \{front, back\}, (w,g) \in \Idx_e \right\} \notag
\end{align}

Now, Molly can compute the data flow between the statements, or def-use chains. For brevity, the exact set constraints are omitted here, but of course there can only be a flow between instances that are in the iteration domain $\Dom$.

Equations~\ref{eqf:fieldflow} to~\ref{eql:fieldflow} describe the direct data flow between the statements. These are data transfers within a SCoP whose chunk size will be computed later. Equations~\ref{eqf:inputflow} to~\ref{eql:inputflow} are values that must be read from the local storage and potentially transferred to another node if the consumer statement is executed on another node. The data flows~\ref{eqf:outputflow} to~\ref{eql:outputflow} are not overwritten by any other statement and therefore must be written back to the local storages.
 
\newcommand*{\flow}{\;\delta\;}
Field flow deps:
\begin{align}
  (S1.7,(i,x,y)) &\flow (S2.1,(i,x,y)) \label{eqf:fieldflow} \\
  (S2.2,(i-1,x-1,y)) &\flow (S1.1,(i,x,y)) \label{eqf:fieldflowex}\\
  (S2.2,(i-1,x,y+1)) &\flow (S1.2,(i,x,y))\\
  (S2.2,(i-1,x+1,y)) &\flow (S1.3,(i,x,y))\\
  (S2.2,(i-1,x,y-1)) &\flow (S1.4,(i,x,y)) \label{eql:fieldflow} \\
  (Prologue,()) &\flow (S1.1,(0,x,y)) \label{eqf:inputflow}\\
  (Prologue,()) &\flow (S1.2,(0,x,y))\\
  (Prologue,()) &\flow (S1.3,(0,x,y))\\
  (Prologue,()) &\flow (S1.4,(0,x,y))\\
  (Prologue,()) &\flow (S1.5,(0,x,y)) \label{eql:inputflow}\\
  (S1.7,(99,x,y)) &\flow (Epilogue,()) \label{eqf:outputflow}\\
  (S2.2,(99,x,y)) &\flow (Epilogue,()) \label{eql:outputflow}
\end{align}

Equations~\ref{eqf:nonfieldflow} to~\ref{eql:nonfieldflow} are non-field flows, i.e. they are stored and loaded from memory that is always local. Therefore, they force  related instances to be executed on the same node.

\begin{align}
  (S1.1,(i,x,y)) &\flow (S1.2,(i,x,y)) \label{eqf:nonfieldflow}\\
  (S1.2,(i,x,y)) &\flow (S1.3,(i,x,y))\\
  (S1.3,(i,x,y)) &\flow (S1.4,(i,x,y))\\
  (S1.4,(i,x,y)) &\flow (S1.6,(i,x,y))\\
  (S1.5,(i,x,y)) &\flow (S1.6,(i,x,y))\\
  (S2.1,(i,x,y)) &\flow (S2.2,(i,x,y)) \label{eql:nonfieldflow}
\end{align}

The geometry of the nodes must be fixed here. In this example, the cluster consists of 64 nodes in an $8 \times 8$ mesh. The block distribution result in $128 \times 128$-sized tiles.

\begin{align}
  \Proc &= \{ (a,b) \mid 0 \le a,b < 8 \} \notag\\
  \pi_{front,back}(w,h) &= (\lfloor w/128 \rfloor,\lfloor h/128 \rfloor) \label{eq:pifield}
\end{align}

Based on the owner-computes policy, one possible distribution of statement instances to executions is presented below.

\begin{align*}
  \pi_{S1.7}(i,x,y) &= \pi_{back}(x,y)\\
  \pi_{S2.2}(i,x,y) &= \pi_{front}(x,y)
\end{align*}

The non-field dependencies~\ref{eqf:nonfieldflow} to~\ref{eql:nonfieldflow} now force the rest of the statements to be computed on the same nodes as the consumers:

\begin{align*}
  \pi_{S1.1 \dots S1.6}(i,x,y) &= \pi_{S1.7}(i,x,y) = \pi_{back}(x,y)\\
  \pi_{S1.2}(i,x,y) &= \pi_{S2.2}(i,x,y) = \pi_{front}(x,y)
\end{align*}

Next, one has to find a chunking function for the intra-SCoP flow dependencies. This example shows the chunking of one of the flows, the one of Equation~\ref{eqf:fieldflowex}. The Algorithm~\ref{alg:chunking} computes a level of independence of 2 since the for-loop on $i$ serializes the two interior loops. In addition, no dependence is violated. The chunking function therefore is $\varphi_{S1.1}(i,x,y) = (S1.1, (i,1,1))$, or any other representative consumer with fixed $i$.

The transfer relation therefore becomes the Equation~\ref{eq:transfers}. Since the generator instances S2.2 are each executed on exactly one node, no unique producer has to be chosen and $Transfers' = Transfers$. The chunks become Equation~\ref{eq:chunks}.

\begin{figure*}
\begin{align}
Transfers &= \big\{ (S1.1, (i,1,1), S2.2, (i-1,x-1,y), (\lfloor (x-1)/128 \rfloor, \lfloor y/128 \rfloor), S1.1, (i,x,y), (\lfloor x/128 \rfloor, \lfloor y/128 \rfloor)) \big\} \label{eq:transfers} \\
  Chunks &= \big\{  \{ ( \underbrace{S2.2, (i-1,x-1,y), \pi_{front}(x-1,y)}_\text{producer}, \underbrace{S1.1, (i,x,y), \pi_{front}(x,y))}_\text{consumer}, \underbrace{front, (x,y)}_\text{field element} \}_{0 \le i < 99} \big\} \label{eq:chunks}
\end{align}
\end{figure*}

For transferring the data between statements $S2.2$ and $S1.1$ two possibilities remain: The case $\pi_{front}(x-1,y) = \pi_{front}(x,y)$, i.e. the value is used on the same node where it is computed, and $\pi_{front}(x-1,y) \neq \pi_{front}(x,y)$ i.e. the value is computed on on the neighbor node. These are the elements $ \{ (front, (w,h)) \in \Idx_{front} \mid  w = 128a - 1 , 128b \le h < 128(b+1) \}$ that need to be transferred from node $(a-1,b)$ (where the producer executes) to $(a,b)$ (where the consumer executes). Per source node/target node either 128 or 127 (at the extremities) values have to be transferred. A possible indexing function into the communication buffer can be $f(w,h) = h - 128b$, where $b = \lfloor h/128 \rfloor$ as of Equation~\ref{eq:pifield}.

Therefore, on the producer node, Molly generates three statements to prepare the send buffer, write the value into the send buffer and then start the transfer. Analogously, there are three statements on the receiver side. For instance, Molly generates a statement $S_\text{recv}$ on the receiving node $(a,b)$ that calls \texttt{\_\_molly\_recv}($(a-1,b)$) with the properties

\begin{align*}
  \Dom_\text{recv} &= \{ (i) \mid 1 \le i < 100  \}\\
  \theta_\text{recv}(i) &= (0,i,128a,0,128b,-0.5)\\
  \pi_\text{recv}(i) &= \{ (a,b) \in \Proc \}
\end{align*}

\section{Experiments}\label{sct:experiments}

The example program in Figure~\ref{fig:conway} has been executed on the Blue Gene/Q supercomputer in Jülich. The results are shown in Table~\ref{tab:results}, with varying field sizes (1024 in the Figure) and on different number of nodes. The times are averaged over three executions of the SCoP with one warmup execution before.

\begin{table}
\begin{tabular}{llllll}
Geometry & RpN\footnotemark[3] & Field size & TpN\footnotemark[4] & Time & MegaLUPS\footnotemark[5] \\
\hline
2x2           & 1              & $256^2$  & $128^2$ &  0.9 s  & 7 \\ 
4x4           & 1              & $512^2$  & $128^2$ &  1 s   & 26.1 \\ 
8x8           & 1              & $512^2$  & $64^2$ &  0.3 s  & 98.1 \\ 
8x8           & 1              & $1024^2$  &  $128^2$ &  1   s & 104.8 \\ 
8x8           & 1              & $2048^2$  &  $256^2$ &  4   s & 104.5  \\ 
8x8           & 1              & $4096^2$  & $512^2$ &  15.8 s  & 106.3 \\ 
8x8           & 1              & $8192^2$  & $1024^2$ &  63 s  & 106.5 \\ 
16x16         & 1              & $2048^2$  & $128^2$ &  1 s  & 418.9 \\ 
32x32         & 1              & $4096^2$  & $128^2$ &  1 s  & 1577.2 \\ 
32x32         & 16              & $4096^2$  & $128^2$ &  1 s  & 1669.8 \\ 
32x32         & 64              & $4096^2$  & $128^2$ &  1.7 s  & 1010 \\ 
64x64         & 64              & $8192^2$  & $128^2$ &  1.7 s  & 4064.6 \\ 
128x128       & 64              & $16384^2$  & $128^2$ &  1.7 s  & 16168.6 \\ 
\end{tabular}
\centering
\caption{Runtimes of the reduced Game of Life (Figure~\ref{fig:conway}) on Blue Gene/Q.}\label{tab:results}
\end{table}
\footnotetext[3]{Ranks per node (virtual MPI nodes per physical processor)}
\footnotetext[4]{Tile size per rank}
\footnotetext[5]{Million lattice-site updates per second (executions of \texttt{hasLife})}

The experiments show the approximately perfect weak scaling of the program and on this hardware. If the tile size per node is $128^2$, the 100 iterations take almost exactly one second, whatever the total number of nodes in the system. The nodes on the mesh surface have less work to do, therefore the execution with just 4 nodes takes just 0.9 seconds.

The Blue Gene/Q compute nodes have 16 physical cores with 4-way SMT\footnote{Hardware threads} each. Up to 64 ranks can be put onto one node, in which case the execution time increases to just 1.7 seconds, but gets even a little faster with one rank per core due to in-memory transfers on the node.

The program also scales as expected if the size of the tile per rank increases. Per 4-times larger time, the program takes 4 times longer to execute. The number of stencil executions per seconds remains approximately the same.

Of course, this serves more as a demonstration of the capabilities of Molly than to run the Game of Life. Any optimized Game of Life program reaches this performance without thousands of nodes with techniques that are not in Molly's scope. In addition, the code generated by Molly is not as optimal as it could. For instance, it generates a lot of unnecessary conversions between 32- and 64-bit integers.

\section{Extensions}\label{sct:extensions}

The current implementation is fairly limiting and yet unusable for practical uses. In the current state Molly is no more than a prototype. The most severe missing capabilities are

\begin{itemize}
\item Reductions are not supported. In the current implementation a reduction is detected as a non-field dependency that cases all statements to be executed on every node.
\item Field elements must be scalars. As pointer arithmetic on element pointers are forbidden, no byte offset can be added before reading or writing to any member except the first.
\item Support for periodic boundary conditions (arrays that are tori instead of meshes)
\item Runtime information on the chosen data layout. This is necessary, for instance, to implement filling a field with data from a file, where each node loads a section of the file and stores it locally.
\item Fields must be global variables. Fields local to a function or as a struct member do not work. Also, fields cannot be passed as arguments to functions.
\item Conditionals inside the SCoP, even if statically determined, are unsupported. Conditionals on affine constraints can just be subtracted from the iteration domain. Other conditions might be handled as a no-op write of the condition is not met (F[$\vec k$] = cond ? newValue : F[$\vec k$]).
\end{itemize}

At least the first four capabilities are required for LatticeQCD simulation and therefore planned as part of my PhD thesis. Some more visionary features are presented in the following.

\subsection{Explicit Data Distribution}

The current implementation always uses a fixed block-distribution with equally sized blocks of data for every node. Data regions do not overlap. Further development will allows the programmer to define an arbitrary mapping by annotation the field declaration with a \texttt{\#pragma}. For instance, the following annotation interchanges the index dimensions such that the data is stored in column-major format, instead of the default of row-major in C, by reversing the indexing. The home location(s) can be selected as well.

\begin{verbatim}
#pragma molly transform("{ [i,j] -> [j,i] }", 0)
molly::array<bool,128> field;
\end{verbatim}

Further ideas include tiling of memory areas, grouping elements for access using SIMD-style loads, using space-filling curves (Hilbert Curve, Z-Curve) for local indexing and compiler-generated indirect addressing (F[Idx[$\vec k$]]) if the index function becomes too complicated. One might even switch the data layout dynamically at runtime based on what the next operation on it will be\footnote{The author did this for the manual optimization of lattice QCD}.

Furthermore, Molly might decide itself where to place data based on heuristics on the SCoPs that access them.

\subsection{Explicit Statement Distribution}

Instead of employing a policy like \dquote{owner computes}, the user might explicitly state where to execute a statement. This gives more control to the programmer who might know better what gives the best performance or wants to try out different possibilities.

{\small
\begin{verbatim}
#pragma molly where("{[i,x,y]->[floor(x/128),floor(y/128)]}")
auto living = hasLife(front[x][y], neighbors);
\end{verbatim}
}


\section{Conclusion}\label{sct:conclusion}

Molly is a compiler extension intended to take away the complexity of writing scientific programs for distributed memory machines by generating the necessary communication code. Such communication code often has the same patterns but is difficult to write because a lot of index computations, allocations, node synchronizations etc. are involved. The goal is not to take away control from the programmer who can still decide where to place and execute computations and data. Quite the contrary, the developer can try out different configurations to see which runs faster on a particular machine.

In contrast to previous attempts on automatic parallelization on distributed memory machines Molly does not try to be the one tool that fits it all but optimizes only special cases well. Nor does it introduce a new programming language that forces the developer to only use high-level constructs. The programmer does not need to write explicitly for a specific architecture but still has full control if needed for optimization. The integration into the general purpose compiler Clang+LLVM hopefully simplifies its use. It also has the advantage of taking away the necessity of reinventing the wheel for every language feature or optimization.

The first experiments give a promising outlook on what is possible for a large class of scientific programs. These include stencil-computations, image processing, linear algebra, differential equations, quantum chemistry, quantum physics and generally anything that can be expressed as a static control part (SCoP). It is not intended for other things like bioinformatics, data(base) lookups, or graphs computations.

This work will be part of the author's PhD thesis that aims to automatically parallelize Lattice QCD programs for various types of cluster architectures. It grew from the author's experience of optimizing such a program manually with many different optimization where most involved writing just a lot of boilerplate code. A comparison of the manually and Molly-optimized program will appear in the PhD thesis.

\acks

\emph{Not yet written}

\bibliographystyle{abbrvnat}

\begin{thebibliography}{0}
\providecommand{\natexlab}[1]{#1}
\providecommand{\url}[1]{\texttt{#1}}
\expandafter\ifx\csname urlstyle\endcsname\relax
  \providecommand{\doi}[1]{doi: #1}\else
  \providecommand{\doi}{doi: \begingroup \urlstyle{rm}\Url}\fi

\end{thebibliography}


\begin{thebibliography}
\softraggedright


\bibitem{tmlqcd}
K. Jansen, C. Urbach. \emph{tmLQCD: A Program suite to simulate Wilson Twisted mass Lattice QCD}. May 2009. 44 pp.
Published in Comput.Phys.Commun. 180 (2009) 2717-2738
arXiv:0905.3331 [hep-lat]

\bibitem{bonati13}
C. Bonati, M. D'Elia, M. Mariti, F. Negro and F. Sanfilippo. \emph{Equation of state and magnetic susceptibility of $N_f = 2 + 1$ QCD with physical quark masses}. arXiv:1310.8656 [hep-lat]




\bibitem{fautrier92a}
P. Feautrier. \emph{Some efficient solutions to the affine scheduling problem. I. One-dimensional time}. International journal of parallel programming 21, no. 5 (1992): 313-347.

\bibitem{fautrier92b}
P. Fautrier. \emph{Some efficient solutions to the affine scheduling problem. Part II. Multidimensional time}. In: International Journal of Parallel Programming 21 (6 1992), pp. 389–420.

\bibitem{darte00}
A. Darte, Y. P. Robert, Frederic Vivien, et al. \emph{Scheduling and automatic Parallelization}. Springer, 2000.


\bibitem{mauras90}
C. Mauras, P. Quinton, S. Rajopadhye and Y. Saouter. \emph{Scheduling affine parameterized recurrences by means of variable dependent timing functions}. In Proceedings of the International Conference on Application Specific Array Processing, 1990. pp. 100-110. IEEE, 1990.

\bibitem{quillere02}
F. Quilleré, and S, Rajopadhye. \emph{Optimizing memory usage in the polyhedral model}. ACM Transactions on Programming Languages and Systems (TOPLAS) 22, no. 5 (2000): 773-815.

\bibitem{grosslinger09}
A. Gr{\"o}{\ss}linger. \emph{Precise management of scratchpad memories for localising array accesses in scientific codes}. In Compiler Construction, pp. 236-250. Springer Berlin Heidelberg, 2009.


\bibitem{llvm}
C. Lattner and V. Adve. \emph{LLVM: A compilation framework for lifelong program analysis \& transformation}. In Code Generation and Optimization, 2004. CGO 2004. International Symposium on, pp. 75-86. IEEE, 2004.

\bibitem{clang}
C. Lattner et. al. \emph{Clang: A C Language Family Frontend for LLVM}. {http://clang.llvm.org} 



\bibitem{polly}
T. Grosser, A, Gr{\"o}{\ss}linger and C. Lengauer. \emph{Polly - Performing polyhedral optimizations on a low-level intermediate representation}. Parallel Processing Letters 22, no. 04 (2012).

\bibitem{isl}
S. Verdoolaege. \emph{isl: An integer set library for the polyhedral model}. In Mathematical Software–ICMS 2010, pp. 299-302. Springer Berlin Heidelberg, 2010.


\bibitem{kennedy07}
K. Kennedy, C. Koelbel and H. Zima. \emph{The rise and fall of High Performance
Fortran: an historical object lesson}. In: Proceedings of the third ACM SIGPLAN
conference on History of programming languages. HOPL III. San Diego, California:
ACM, 2007, pp. 7–1–7–22

\bibitem{adve98}
V. Adve and J. Mellor-Crummey. \emph{Using integer sets for data-parallel program analysis and optimization}. In ACM SIGPLAN Notices, vol. 33, no. 5, pp. 186-198. ACM, 1998.

\bibitem{omega}
W. Kelly, V. Maslov, W. Pugh, E. Rosser, T. Shpeisman and D. Wonnacott. \emph{The Omega calculator and library, version 1.1.0.} College Park, MD 20742 (1996): 18.

\bibitem{amarasinghe93}
S. P. Amarasinghe and M. S. Lam. \emph{Communication optimization and code generation for distributed memory machines}. In ACM SIGPLAN Notices, vol. 28, no. 6, pp. 126-138. ACM, 1993.

\bibitem{dathathri13}
R. Dathathri, C. Reddy, T. Ramashekar, and U. Bondhugula. \emph{Generating efficient data movement code for heterogeneous architectures with distributed-memory}. In Parallel Architectures and Compilation Techniques (PACT), 2013 22nd International Conference on, pp. 375-386. IEEE, 2013.





\bibitem{bondhugula11}
U. Bondhugula. \emph{Automatic Distributed-Memory Parallelization and Code Generation
using the Polyhedral Framework}. IISc-CSA-TR-2011-3. Bangalore: Indian Institute
of Science, Sept. 2011.

\bibitem{classen06}
M. Classen and M. Griebl. \emph{Automatic code generation for distributed memory architectures in the polytope model}. In Parallel and Distributed Processing Symposium, 2006. IPDPS 2006. 20th International. IEEE, 2006. p. 7 pp.

\bibitem{yuki13}
T. Yuki and S. Rajopadhye. \emph{Memory Allocations for Tiled Uniform Dependence Programs}. IMPACT 2013 (2013): 13.



\bibitem{ravishankar12}
M. Ravishankar, J. Eisenlohr, L.-N. Pouchet, J. Ramanujam, A. Rountev and P. Sadayappan. \emph{Code generation for parallel execution of a class of irregular loops on distributed memory systems}. In Proceedings of the International Conference on High Performance Computing, Networking, Storage and Analysis, p. 72. IEEE Computer Society Press, 2012.


\bibitem{snyder99}
L. Snyder. \emph{A Programming Guide to ZPL}. MIT Press, Scientific and Engineering Computation Series, March 1999

\bibitem{bastoul03}
C. Bastoul and P. Feautrier. \emph{Improving data locality by chunking. In CC'12 International Conference on Compiler Construction}. LNCS 2622, pages 320-335, Warsaw, Poland, Apr. 2003

\bibitem{pluto}
U. Bondhugula, A. Hartono, J. Ramanujam and P. Sadayappan. \emph{A practical automatic polyhedral parallelizer and locality optimizer}. In ACM SIGPLAN Notices, vol. 43, no. 6, pp. 101-113. ACM, 2008.


\bibitem{cloog}
C. Bastoul. \emph{Code generation in the polyhedral model is easier than you think}. In Proceedings of the 13th International Conference on Parallel Architectures and Compilation Techniques, pp. 7-16. IEEE Computer Society, 2004.


\end{thebibliography}

\end{document}